\documentclass[12pt]{article}
\usepackage{latexsym,amssymb,amsmath}

\begin{document}

\title
{Spontaneous compactification and nonassociativity}
\author
{E.K. Loginov\footnote{{\it E-mail address:} ek.loginov@mail.ru}\\
\it Department of Physics, Ivanovo State University\\
\it Ermaka St. 39, Ivanovo, 153025, Russia}
\date{}
\maketitle

\begin{abstract}
We consider the Freund-Rubin-Englert mechanism of compactification of $N=1$ supergravity in 11
dimensions. We systematically investigate both well-known and some new solutions of the
classical equations of motion in 11 dimensions. In particular, we show that any threeform
potential in 11 dimension is given locally by the structure constants of a geodesic loop in an
affinely connected space.
\end{abstract}

\section{Introduction}

The Kaluza-Klein mechanism of spontaneous compactification works as follows~[1-4]. The
equations describing gravity and matter fields in $d$ dimensions are considered. A vacuum
solution of the equations in which $d$-dimensional spacetime $M$ is of the form $M_4\times K$
is searched. Here $M_4$ is a maximally symmetric four-dimensional space (de Sitter space,
anti--de Sitter space or Minkowski space) and $K$ is a compact manifold (as a rule this is an
Einstein space). The representation of $M$ in the form of a direct product induces the block
diagonal form of the vacuum metric
\begin{equation}\label{51-13}
g_{MN}=\begin{pmatrix} g_{\mu\nu}&0\\0&g_{mn}\end{pmatrix},
\end{equation}
where $g_{\mu\nu}$ and $g_{mn}$ are components of $g_{MN}$ defined on $M_4$ and $K$
respectively. In one's turn, the block representation of $g_{MN}$ is compatible with the
Einstein equations
\begin{equation}
R_{MN}-\frac12g_{MN}R=T_{MN}-\Lambda g_{MN},
\end{equation}
if components of the energy-momentum tensor of matter fields are
\begin{equation}\label{51-01}
\begin{aligned}
T_{\mu\nu}&=k_1g_{\mu\nu},\\
T_{mn}&=k_2g_{mn}.
\end{aligned}
\end{equation}
If we consider the interaction of gravity with matter fields without the potential term
$g_{MN}V(\varphi)$, then $T_{00}>0$ and the constant $k_1$ is negative. In addition, if the
cosmological constant $\Lambda=0$, then it follows from (\ref{51-01}) that $M_4$ is an
anti--de Sitter space. On the contrary, if $T_{MN}$ contains the potential term, then the
solution $\varphi=\text{const}$ is equivalent to the introduction of the $\Lambda$ term and
the space $M_4$ may be flat~\cite{marc82,nish84}.
\par
Now we briefly consider the Freund-Rubin-Englert mechanism~\cite{freu80,engl82} of spontaneous
compactification of $d=11$ supergravity. In the Bose sector of this theory the equations of
motion (Einstein equations and equations for the antisymmetric gauge field strength) have the
form~\cite{crem78}
\begin{align}
R_{MN}-\frac12g_{MN}R&=12\left(8F_{MPQR}F_{N}{}^{PQR}-g_{MN}F_{SPQR}F^{SPQR}\right),
\label{51-11}\\
F^{MNPQ}{}_{;M}&=-\frac{\sqrt2}{24}\varepsilon^{NPQM_1\dots
M_8}F_{M_1M_2M_3M_4}F_{M_5M_6M_7M_8}\label{51-12},
\end{align}
where $\varepsilon^{M_1\dots M_{r}}$ is a fully antisymmetric covariant constant tensor such
that $\varepsilon_{1\dots r}=|g|^{1/2}$. The Freund and Rubin solution~\cite{freu80} is
\begin{equation}\label{51-21}
F_{\mu\nu\sigma\lambda}=\rho\varepsilon_{\mu\nu\sigma\lambda},
\end{equation}
where $\rho$ is a real constant, and all other components of $F_{MNPQ}$ are zero. A more
composite solution is obtained if we assume that $F_{MNPQ}\ne0$ not only for the spacetime
components assuming the Freund-Rubin form, but also for the internal space components. Such
solutions (Englert solution~\cite{engl82}) were first constructed on the sphere $\mathbb S^7$
with torsion. The Englert solution is to set
\begin{align}
F_{\mu\nu\sigma\lambda}&=\rho\varepsilon_{\mu\nu\sigma\lambda},\\
F_{mnpq}&=\lambda\partial_{[q}S_{mnp]}\label{51-02},
\end{align}
where $S_{mnp}=S_{[mnp]}$ is a suitable totally antisymmetric torsion tensor.
\par
Note that the connection between an antisymmetric gauge field strength and a torsion defined
by (\ref{51-02}) has an universal character in the 11-dimensional supergravity. Bars and
McDowell~\cite{bars83} have shown that the $g_{MN}/A_{MNP}$ gravity-matter system may be
reinterpreted, in first-order formalism, as a pure gravity theory with torsion $S_{MNP}$ such
that
\begin{align}
A_{MNP}&=\lambda S_{[MNP]},\label{51-22}\\
F_{MNPS}&=\partial_{[S}A_{MNP]}.\label{51-23}
\end{align}
In this sense, even the Abelian gauge invariance of $A_{MNP}$ may be regarded as a spacetime
symmetry of pure gravity. Besides, the deformation
\begin{equation}
\overset\circ\Gamma_{MNP}\to\overset\circ\Gamma_{MNP}+S_{[MNP]}
\end{equation}
of the Riemann connection converts $M$ into an affinely connected space with torsion. Note
that such an interpretation is possible only for three-index fields since the torsion tensor
is rank 3.

\section{Preliminaries}

Out next goal is to think about algebraic properties of the threeform potential and its field
strength. The requisite concepts are geodesic loops in an affinely connected space and its
tangent algebras. In~\cite{kikk64} Kikkawa introduced the idea of a geodesic loop in an
affinely connected space. In that article it is proved that in a neighborhood of each point of
an affinely connected space we can define, in a natural way, the operation of multiplication,
relative to which this neighborhood becomes a local loop. Let us recall Kikkawa's
construction.
\par
Let $M$ be an affinely connected space and $e$ an arbitrary point in it. In a neighborhood of
this point, we introduce a binary operation as follows. Let $x$ and $y$ be two points
belonging to this neighborhood. We connect them to the point $e$ by geodesic lines
$\overset\frown{ex}$ and $\overset\frown{ey}$. Then we translate the arc $\overset\frown{ex}$
in a parallel way into the position $\overset\frown{yz}$. The point $z$ is then by definition
the product of the points $x$ and $y$, i.e. $z=xy$. We denote the obtained binary system by
the symbol $G_{e}$. It is easy to prove that $e$ is the unity element of $G_{e}$, the
equations $ax=b$ and $ya=b$ are uniquely solvable for all $a,b\in G_{e}$, and $G_{e}$
satisfies the identity
\begin{equation}\label{51-05}
xx^2=x^2x.
\end{equation}
Thus, $G_{e}$ is a local monoassociative loop. The loop $G_{e}$ is said to be a geodesic loop
of the affinely connected space $M$.
\par
To study geodesic loops in affinely connected spaces we may use a method which is usually
employed to study the local structure of Lie groups. Despite the lack of associativity in
geodesic loops, this method enables us to uniquely define binary $[x,y]$ and ternary $(x,y,z)$
operations in their tangent spaces $T_{e}$ and to construct the local
algebras~\cite{akiv76,akiv06}. These operations are expressed in terms of the coordinates of
the vectors $x,y,z\in T_{e}$ as follows:
\begin{align}
{}[x,y]^{i}&=2\alpha^{i}_{jk}x^{j}y^{k},\label{51-03}\\
(x,y,z)^{i}&=2\beta^{i}_{jkl}x^{j}y^{k}z^{l},\label{51-04}
\end{align}
If $\alpha^{i}_{jk}=0$, then the corresponding geodesic loop $G_{e}$ is Abelian, if
$\beta^{i}_{jkl}=0$, then the loop $G_{e}$ is associative, i.e. it is a local Lie group. The
tensors $\alpha^{i}_{jk}$ and $\beta^{i}_{jkl}$ are called the fundamental tensors of the
geodesic loop $G_{e}$~\cite{akiv78}. They satisfy the relation
\begin{equation}\label{51-10}
\beta^{i}_{[jkl]}=\alpha^{m}_{[jk}\alpha^{i}_{l]m}
\end{equation}
which follows from (\ref{51-05}). The right-hand side of the relation is obtained from the
Jacobiator of the vectors $x$, $y$ and $z$. This is the reason that relation (\ref{51-10}) is
called the generalized Jacobi identity.
\par
Now suppose $M$ is an affinely connected space with torsion and $\Gamma^{i}_{jk}$ is a
metric-compatible affine connection, the difference with the Levi-Civita connection is given
by the contortion tensor. It is easy to prove that $S_{ijk}=S_{[ijk]}$. Therefore
\begin{equation}\label{51-14}
\Gamma_{ijk}=\overset\circ\Gamma_{ijk}+S_{ijk}
\end{equation}
is a metric-compatible affine connection with skew-symmetric torsion. We choose the torsion
and curvature tensors of $M$ in the form
\begin{align}
S^{i}_{jk}&=\Gamma^{i}_{[jk]},\\
R^{i}_{jkl}&=\partial_{k}\Gamma^{i}_{jl}-\partial_{l}\Gamma^{i}_{jk}
+\Gamma^{m}_{jl}\Gamma^{i}_{mk}-\Gamma^{m}_{jk}\Gamma^{i}_{ml}.\label{51-60}
\end{align}
It is known ~\cite{akiv78} that for any geodesic loop constructed in a neighborhood of $e\in
M$, the fundamental tensors can be expressed using values of the torsion and curvature tensors
in $e$ by the formulas
\begin{align}
\alpha^{i}_{jk}&=-S^{i}_{jk},\label{51-07}\\
4\beta^{i}_{jkl}&=-2\nabla_{l}S^{i}_{jk}-R^{i}_{jkl}.\label{51-08}
\end{align}
Thus, noncommutativity and nonassociativity of the local geodesic loops are intimately related
to torsion and curvature of the space $M$. Note that the formulas (\ref{51-07}) and
(\ref{51-08}), fully antisymmetry of the torsion tensor, and the Bianchi identities
\begin{align}
R^{i}_{[jkl]}+2\nabla_{[j}S^{i}_{kl]}+4S^{m}_{[jk}S^{i}_{l]m}&=0,\label{51-09}\\
\nabla_{[k}R^{ij}{}_{lm]}-2R^{ij}{}_{n[k}S^{n}_{lm]}&=0\label{51-20}
\end{align}
will play an important part in following constructions. Note also that the first Bianchi
identity (\ref{51-09}) is obtained if we substitute (\ref{51-07}) and (\ref{51-08}) in the
identity (\ref{51-10}).

\section{The Freund-Rubin type solutions}

In this section, we begin an analysis of $M_4\times K$ compactification of $d=11$
supergravity. We consider the Bose sector of this theory as a pure gravity theory with
torsion. We show that the torsion is given locally by structure constants of a geodesic loop.
Then we apply this result to an analysis of the Freund-Rubin solution.

\subsection{Geodesic groups}

We consider the $M=M_4\times K$ compactification of $d=11$ supergravity. Obviously, $M$ is a
Riemann space with a metric of the block diagonal form (\ref{51-13}). We deform the Riemannian
connection by the rule (\ref{51-14}) and convert $M$ into an affinely connected space with a
fully antisymmetric torsion tensor $S_{ijk}$. Since the projections $M_4\times K\to M_4$ and
$M_4\times K\to K$ are Riemannian submersions, we can construct objects in $M$ by pulling back
the objects from $M_4$ and $K$ along these projections. In particular, we consider geodesic
loops in $M_4$ and in $K$ separately in order to later pull them back to $M$ in this way.
\par
Suppose $e$ is an arbitrary point in $M_4$ or $K$. In a neighborhood of this point, we define
a geodesic loop $G_{e}$. Then using the full skew-symmetry of $S_{ijk}$, we rewrite the
tensors (\ref{51-07}) and (\ref{51-08}) in the form
\begin{align}
\alpha_{ijk}&=-S_{ijk},\label{51-18}\\
4\beta_{ijkl}&=-2S_{ijk;l}+6S^{m}_{[ij}S_{kl]m}-R_{ijkl}\label{51-19}.
\end{align}
Now suppose the loop $G_{e}$ is associative. Then we have the identities
\begin{align}
\beta_{ijkl}&=0,\\
\alpha^{m}_{[jk}\alpha^{i}_{l]m}&=0\label{51-16}
\end{align}
instead of (\ref{51-10}). It follows from here that the tangent algebra $A_{G}$ of the local
loop $G_{e}$ in the point $e$ is a Lie algebra. We prove that this algebra is compact if $e\in
K$. Indeed, we choose the basic $\{e_{i}\}$ in $A_{G}$ such that
\begin{equation}\label{51-68}
[e_{i},e_{j}]=\alpha_{ij}{}^{k}e_{k}
\end{equation}
and define the bilinear form
\begin{equation}\label{51-67}
(e_{i},e_{j})=g_{ij}.
\end{equation}
Using the full skew-symmetry of $\alpha_{ijk}$, we prove the equality
\begin{equation}\label{51-56}
([e_{i},e_{j}],e_{k})=(e_{i},[e_{j},e_{k}]).
\end{equation}
If the point $e\in K$, then the inner product (\ref{51-67}) is Euclidean and hence it is
positive definite. Therefore the Lie algebra $A_{G}$ is compact. It is well known (see
e.g.~\cite{jaco62}) that any compact Lie algebra is reductive, and hence isomorphic to the
direct sum of a semisimple Lie subalgebra and the center, which is Abelian. Therefore if the
algebra $A_{G}$ is non-Abelian, then we have one of the following isomorphisms:
\begin{equation}\label{51-26}
\begin{aligned}
A_{G}&\simeq su(2)\oplus su(2)\oplus u(1),\\
A_{G}&\simeq su(2)\oplus u(1)\oplus u(1)\oplus u(1)\oplus u(1).
\end{aligned}
\end{equation}
If the point $e\in M_4$, then the inner product (\ref{51-67}) is Lorentzian. The
classification of Lorentzian metric Lie algebras is known~\cite{medi83}. Any such algebra is
isomorphic to one of the following metric Lie algebras:
\par
(1) Abelian with Lorentzian inner product,
\par
(2) $su(2)\oplus u(1)$ with a timelike inner product on $u(1)$,
\par
(3) $su(1,1)\oplus u(1)$ with Lorentzian inner product on $su(1,1)$ and Euclidean on $u(1)$,
\par
(4) the Nappi-Witten solvable Lie algebra~\cite{napp93}.
\par\noindent
We find an explicit form of the Lie brackets (\ref{51-68}) for the above Lie algebras.
Obviously, the algebra $A_{G}$ is Abelian if and only if all $\alpha_{ijk}=0$. Suppose 1, 2,
and 3 are spatial indexes. Then
\begin{equation}\label{51-59}
\begin{aligned}
A_{G}&\simeq su(2)\oplus u(1)\\
A_{G}&\simeq su(1,1)\oplus u(1)
\end{aligned}\qquad
\begin{aligned}
&\text{if only}\quad\alpha_{123}\ne0,\\
&\text{if only}\quad\alpha_{124}\ne0.
\end{aligned}
\end{equation}
In order that to find the Lie brackets for the Nappi-Witten Lie algebra, we note that the
algebra has the following explicit description:
\begin{equation}\label{51-69}
[J,P_{i}]=\varepsilon_{ij}P_{j},\quad [P_{i},P_{j}]=\varepsilon_{ij}T,\quad [T,J]=[T,P_{i}]=0.
\end{equation}
This algebra is a central extension of the $d=2$ Poincare algebra to which it reduces if one
sets $T=0$. We suppose
\begin{equation}\label{51-43}
e_{i}=P_{i},\quad e_3=J-T,\quad e_4=J.
\end{equation}
Then it is easily shown that
\begin{equation}\label{51-70}
\alpha_{123}=\alpha_{124}=1,\quad\alpha_{134}=\alpha_{234}=0.
\end{equation}
Thus, if the group $G_{e}$ is non-Abelian, then any nonzero torsion defined in the point $e\in
K$ or $M_4$ is given by structure constants of the algebras (\ref{51-26}) or (\ref{51-59}) and
(\ref{51-69}) respectively. It follows from (\ref{51-22}) that it is true for any threeform
potential with components that take nonzero expectation values in $M_4$ or $K$.
\par
Now, we consider the identities (\ref{51-18}) and (\ref{51-19}). Since the curvature tensor
$R_{ijkl}$ is skew-symmetric in the last two indexes and the torsion tensor is fully
antisymmetric, it follows from (\ref{51-19}) that
\begin{equation}\label{51-24}
-R_{ijkl}=2S_{ijk;l}=2\partial_{[l}S_{ijk]}.
\end{equation}
Hence the tensor $R_{ijkl}$ also is fully antisymmetric. We sum the Bianchi identity on the
indexes $i,m$ and use the antisymmetry of $S_{ijk}$ and $R_{ijkl}$. Then we get
\begin{equation}\label{51-25}
R^{i}_{jkl;i}=S^{i}_{n[j}R^{n}_{kl]i}.
\end{equation}
It follows from the identities (\ref{51-16}) and (\ref{51-24}) that
\begin{equation}
R^{i}_{jkl;i}=-2S^{i}_{n[j}S^{n}_{kl]_;i}=0.
\end{equation}
Since the threeform potential and its gauge field strength are connected with the torsion by
the relations (\ref{51-22}) and (\ref{51-23}), we get
\begin{equation}\label{51-55}
F_{ijkl}{}^{;i}=0.
\end{equation}
Thus, if we consider the Bose sector of $d=11$ supergravity as a pure gravity theory with
torsion, then the gauge field strength must satisfy Eq. (\ref{51-55}).
\par
Further, suppose $e$ is an arbitrary point in $M_4$ and $G_{e}$ is a geodesic group defined in
a neighborhood of $e$. Using (\ref{51-14}), we represent the curvature tensor (\ref{51-60}) in
the form
\begin{equation}\label{51-41}
R^{i}_{jkl}=\overset\circ R{}^{i}_{jkl}+(S^{i}_{jl;k}-S^{i}_{jk;l}
+S^{m}_{jl}S^{i}_{mk}-S^{m}_{jk}S^{i}_{ml}).
\end{equation}
Since the tensor $S_{ijk}$ is fully antisymmetric, it follows from (\ref{51-24}) that the
Ricci tensor
\begin{equation}\label{51-71}
\overset\circ R_{ij}=S^{m}_{ik}S^{k}_{mj}.
\end{equation}
Suppose the tangent Lie algebra $A_{G}$ is non-Abelian. Then it is isomorphic to one of the
algebras (\ref{51-59}) and (\ref{51-69}). Therefore in a suitable local coordinate system the
components of $\alpha_{ijk}$ satisfy to one of the conditions (\ref{51-59}) and (\ref{51-70}).
Then it follows from (\ref{51-71}) that components of the Ricci tensor in the point $e$ have
the form
\begin{equation}
\overset\circ R_{i4}=0,\quad\overset\circ R_{i3}=0\quad\text{or}\quad\overset\circ
R_{i1}=\overset\circ R_{i2}=0.
\end{equation}
Since the metric $g_{ij}$ is nondegenerate, it follows from here that $\overset\circ
R_{ij}\ne\lambda g_{ij}$ if $\lambda$ is nonzero, i.e. the spacetime $M_4$ is non-Einstein.
Now suppose $G_{e}$ is a non-Abelian geodesic group defined in a neighborhood of $e\in K$.
Using the isomorphisms (\ref{51-26}) and arguing as above, we easy prove that the internal
space $K$ is non-Einstein. Thus, if $M=M_4\times K$ is an Einstein space, then the geodesic
loop defined in a neighborhood of any point of $M$ is either nonassociative or it is an
Abelian group. It follows from here that there is no Freund-Rubin background $M_4\times K$ of
11-dimensional supergravity where $M_4$ and $K$ are Lie groups with bi-invariant metric. This,
of course, a known result.

\subsection{The Freund-Rubin ansatz}

Now we start to analyze the Freund-Rubin solution. Suppose $M_4$ is a four-dimensional Riemann
spacetime of signature $(+++-)$. We deform the Riemannian connection by the rule (\ref{51-14})
and convert $M_4$ into an affinely connected space with a fully antisymmetric torsion tensor
$S_{ijk}$. Let $e$ be an arbitrary point in $M_4$. In a neighborhood of this point, we define
a geodesic loop $G_{e}$ and consider the right-hand side of the identity (\ref{51-10}). Since
the tensor $\alpha_{ijk}=g_{is}\alpha^{s}_{jk}$ is fully antisymmetric and its indices take
only four different values, we have
\begin{align}
\alpha^{m}_{[jk}\alpha^{i}_{l]m}&=0,\label{51-57}\\
\beta^{i}_{[jkl]}&=0\label{51-58}
\end{align}
instead of (\ref{51-10}). It follows from (\ref{51-57}) that with respect to the operation of
commutation the tangent algebra $A_{G}$ is a Lie algebra. Suppose the algebra $A_{G}$ is
non-Abelian. Then arguing as above, we see that it has the form (\ref{51-59}) or
(\ref{51-69}). Thus, any nonzero torsion in $e\in M_4$ is given by the structure constants of
these algebras. It follows from (\ref{51-22}) that it is true for any threeform potential with
the components taking nonzero expectation values in $M_4$. Note that the similar assertion for
$S^7$ torsion and the Cayley structure constants was proved in~\cite{gurs83}, where the
Englert's solution~\cite{engl82} was analyzed.
\par
Antisymmetrizing the curvature tensor (\ref{51-41}) on $i$, $j$, and $k$, we get
\begin{equation}\label{51-65}
R_{[ijkl]}=-2S_{ijk;l}=-2\partial_{[l}S_{ijk]}.
\end{equation}
Hence the gauge field strength
\begin{equation}
F_{ijkl}=\lambda' R_{[ijkl]}.
\end{equation}
Substituting $F_{ijkl}$ in the Bianchi identity (\ref{51-20}) and summing on $i$ and $m$, we
obtain
\begin{equation}\label{51-66}
F^{i}_{jkl;i}=S^{i}_{n[j}F^{n}_{kl]i}.
\end{equation}
Since the tensor indices in (\ref{51-66}) take only four different values, the covariant
derivative
\begin{equation}
F_{ijkl;m}=0.
\end{equation}
Hence, the gauge field strength $F_{ijkl}$ is proportional to the fully antisymmetric
covariant constant 4 tensor, i.e. it must have the form (\ref{51-21}). Note that this
assertion is true for any four-dimensional Riemann spacetime of Lorentzian signature. Indeed,
using the full skew-sym\-metry of $S_{ijk}$, the Jacobi identity (\ref{51-57}), and the
equality (\ref{51-41}), we may represent the fundamental tensor (\ref{51-08}) in the form
\begin{equation}
4\beta_{ijkl}=S^{m}_{ij}S_{klm}-\overset\circ R_{ijkl}.
\end{equation}
Since the Riemannian curvature tensor is satisfied the Bianchi identity
\begin{equation}
\overset\circ R{}^{i}_{[jkl]}=0,
\end{equation}
and the conditions (\ref{51-57}) and (\ref{51-58}) do not impose any restrictions on the
spacetime $M_4$.

\section{The Englert's type solutions}

In this section, we continue the investigation of the $M=M_4\times K$ compactification of
$d=11$ supergravity. We again consider the Bose sector of this theory as a pure gravity theory
with torsion. However now we suppose that the matter fields have nonvanishing components in
the internal space $K$. We analyze the Englert's type solutions and prove that the torsion is
given locally by the Cayley structure constants.

\subsection{Geodesic Moufang loops}

We recall~(see i.e.~\cite{logi07}) that the algebra $\mathbb O$ of octonions (Cayley algebra)
is a real linear algebra with the canonical basis $1,e_{1},\dots,e_{7}$ such that
\begin{equation}\label{51-61}
e_{i}e_{j}=-\delta_{ij}+c_{ijk}e_{k},
\end{equation}
where the structure constants $c_{ijk}$ are completely antisymmetric and nonzero and equal to
unity for the seven combinations (or cycles)
$$
(ijk)=(123),(145),(167),(246),(275),(374),(365).
$$
The algebra of octonions is not associative but alternative, i.e. the associator
\begin{equation}
(x,y,z)=(xy)z-x(yz)
\end{equation}
is totally antisymmetric in $x,y,z$. The algebra $\mathbb O$ permits the involution
(antiautomorphism of period two) $x\to\bar x$ such that the elements
\begin{equation}
t(x)=x+\bar x\qquad\text{and}\qquad n(x)=\bar xx
\end{equation}
are in $\mathbb R$. It is easy to prove that the quadratic form $n(x)$ is positive definite
and permits the composition
\begin{equation}
n(xy)=n(x)n(y).
\end{equation}
It is follows from here that the set
\begin{equation}\label{51-62}
\mathbb S=\{x\in\mathbb O\mid n(x)=1\}
\end{equation}
is closed relative to the multiplication in $\mathbb O$ and hence it is an analytic loop. The
loop $\mathbb S$ is the unique, up to isomorphism, analytic compact simple nonassociative
Moufang loop. The tangent algebra of $\mathbb S$ is isomorphic to the seven-dimensional
commutator subalgebra
\begin{equation}\label{51-63}
\mathbb M=\{x\in\mathbb O^{(-)}\mid t(x)=0\}.
\end{equation}
The algebra $\mathbb M$ is the unique compact simple non-Lie Malcev algebra and it satisfies
the identity
\begin{equation}\label{51-64}
[[x,y],z]+[[y,z],x]+[[z,x],y]=6(x,y,z).
\end{equation}
The algebra $\mathbb M$ has the canonical basis $e_{1},\dots,e_{7}$. Using (\ref{51-61}) we
can find the commutators and associators of the basis elements
\begin{align}
[e_{i},e_{j}]&=2c_{ijk}e_{k},\\
(e_{i},e_{j},e_{k})&=2c_{ijkl}e_{l},
\end{align}
where $c_{ijkl}$ is a completely antisymmetric nonzero tensor equal to unity for the seven
combinations
$$
(ijkl)=(4567),(2367),(2345),(1357),(1364),(1265),(1274).
$$
\par
Now let $K$ be a compact seven-dimension Riemann space and $G_{e}$ be a geodesic Moufang loop
defined in a neighborhood of $e\in K$. If the loop $G_{e}$ is associative, then its tangent
algebra is either Abelian or a compact Lie algebra of the form (\ref{51-26}). Suppose $G_{e}$
is a nonassociative Moufang loop. Then arguing as above, we prove that its tangent algebra is
a compact seven-dimension non-Lie Malcev algebra. Since any such algebra is isomorphic to the
algebra (\ref{51-63}), it follows that $G_{e}$ is locally isomorphic to the loop
(\ref{51-62}). Therefore we have the identity
\begin{equation}\label{51-33}
\beta^{i}_{jkl}=\alpha^{m}_{[jk}\alpha^{i}_{l]m},
\end{equation}
instead of (\ref{51-10}). It can easily be checked that this identity is equivalent to the
Malcev identity (\ref{51-64}). On the other hand, it follows from (\ref{51-41}) that
\begin{equation}
\frac12R_{[ijk]l}=S^{m}_{[ij}S_{k]lm}-S_{ijk;l}.
\end{equation}
Hence the tensors $\beta_{ijkl}$ and $S_{ijk;l}$ are fully antisymmetric. Then it follows from
(\ref{51-19}) that the curvature tensor $R_{ijkl}$ is also fully antisymmetric.
\par
Conversely, let $G_{e}$ be a geodesic loop and $R_{ijkl}$ be a fully antisymmetric tensor.
Then
\begin{equation}\label{51-28}
\frac12R_{ijkl}=S^{m}_{[ij}S_{kl]m}-S_{ijk;l}
\end{equation}
and hence the tensor $S_{ijk;l}$ is fully antisymmetric. Again using the identity
(\ref{51-19}), we prove the fully antisymmetry of $\beta_{ijkl}$. Therefore we have the Malcev
identity (\ref{51-33}) instead of (\ref{51-10}) and hence $G_{e}$ is a Moufang loop. Thus, the
geodesic loop $G_{e}$ is Moufang if and only if the curvature tensor $R_{ijkl}$ is fully
antisymmetric. In addition, any non-Abelian geodesic Moufang loop in $K$ is locally isomorphic
to either the Lie group (\ref{51-26}) or the nonassociative Moufang loop (\ref{51-62}).

\subsection{The Englert's ansatz}

We consider the Bose sector of $d=11$ supergravity as a pure gravity theory with torsion and
suppose that the matter fields have nonvanishing components in the internal space $K$. We
suppose that $K$ is an Einstein space and $G_{e}$ is a geodesic Moufang loop defined in a
neighborhood of $e\in K$. As we proved above, $G_{e}$ is nonassociative and hence its tangent
algebra is isomorphic to the algebra $\mathbb M$. We select the basis $\tilde e_1,\dots,\tilde
e_7$ in $\mathbb M$ such that
\begin{align}
[\tilde e_{i},\tilde e_{j}]&=2kc_{ijk}\tilde e_{k},\\
(\tilde e_{i},\tilde e_{j},\tilde e_{k})&=2k^2c_{ijkl}\tilde e_{l},
\end{align}
where $k$ is a real constant. By comparing these equalities with (\ref{51-03}) and
(\ref{51-04}), we get the following relations
\begin{align}
\alpha_{ijk}&=kc_{ijk},\label{51-45}\\
\beta_{ijkl}&=k^2c_{ijkl}.\label{51-46}
\end{align}
It is known~\cite{dund84} that the tensors $c_{ijk}$ and $c_{ijkl}$ are connected by
self-duality relations. For the fundamental tensors of $G_{e}$ these relations are
\begin{align}
\varepsilon^{npqlijk}k\alpha_{ijk}&=6\beta^{npql},\\
\varepsilon^{npqlijk}\beta_{ijkl}&=24k\alpha^{npq}.\label{51-32}
\end{align}
In addition, the following identities are true
\begin{align}
\alpha_{ijm}\alpha^{ijn}&=6k^2\delta^{n}_{m},\label{51-29}\\
\beta_{mijk}\beta^{nijk}&=24k^4\delta^{n}_{m},\label{51-52}\\
\alpha_{im}{}^{j}\alpha_{jn}{}^{k}\alpha_{kp}{}^{i}&=3k^2\alpha_{mnp}.\label{51-30}
\end{align}
\par
Now we substitute the Freund-Rubin ansatz (\ref{51-21}) in the equations of motion
(\ref{51-12}). We obtain
\begin{equation}
F^{mnpq}{}_{;m}=\sqrt2\rho\varepsilon^{npqijkl}F_{ijkl},\label{51-27}
\end{equation}
where $\varepsilon^{npqijkl}$ is the fully antisymmetric covariant constant 7 tensor. Further,
it follows from the Bianchi identity (\ref{51-25}) and the identity (\ref{51-28}) that
\begin{align}
\frac12R_{mnpq}{}^{;m}&=S_{tm[n}S^{t}{}_{pq]}{}^{;m}-S_{mnp;q}{}^{;m}\nonumber\\
&=S^{l}{}_{m[n}S^{mt}{}_{p}S_{q]tl}-S^{l}{}_{m[n}S^{m}{}_{pq];l}.
\end{align}
Using (\ref{51-29}) and (\ref{51-30}), we get
\begin{equation}\label{51-31}
S_{npq;m}{}^{;m}+4k^2S_{npq}=0.
\end{equation}
Moreover, it follows from (\ref{51-28}) that
\begin{equation}
S_{npq;m}=\partial_{[m}S_{npq]}.
\end{equation}
Note that these formulas are true for the geodesic loop of any point in the Einstein space
$K$. Note also that the identity (\ref{51-31}) generalizes the Englert's identity that was
found in the work~\cite{engl82}.
\par
By taking into account the obtained identities, we rewrite Eq. (\ref{51-27}) as
\begin{equation}\label{51-37}
4k^2S^{npq}+\sqrt2\rho\varepsilon^{npqijkl}S_{ijk;l}=0.
\end{equation}
We will find solutions of these equations in the form
\begin{equation}\label{51-36}
S_{mnp;q}=hS_{t[mn}S^{t}{}_{pq]}.
\end{equation}
Such ansatz converts (\ref{51-37}) into the (anti)self-duality equations. In order to obtain a
value of $h$, we find
\begin{equation}\label{51-34}
S_{mnp;q}S^{rnp;q}=h^2\beta_{mnpq}\beta^{rmnpq}=24h^2k^4\delta^{r}_{m}.
\end{equation}
On the other hand, it follows from (\ref{51-31}) that
\begin{equation}\label{51-35}
S_{mnp;q}S^{rnp;q}=-S_{mnp}S^{rnp;q}{}_{;q}=24k^4\delta^{r}_{m}.
\end{equation}
By comparing (\ref{51-34}) and (\ref{51-35}), we get $h=\pm1$. Substituting the ansatz
(\ref{51-36}) in (\ref{51-37}), we obtain the values
\begin{equation}\label{51-40}
k=\pm6\sqrt2\rho.
\end{equation}
Obviously, the obtained solution is self-dual as $h=1$ and anti--self-dual as $h=-1$. Besides,
it follows from (\ref{51-28}) that the curvature tensor
\begin{align}
R_{ijkl}&=0\quad\text{if}\quad h=1,\label{51-38}\\
R_{ijkl}&\ne0\quad\text{if}\quad h=-1.\label{51-39}
\end{align}
Note that the equality (\ref{51-38}) is a necessary and sufficient condition of
parallelizibility of $K$. Precisely this condition was used in~\cite{engl82} for a
construction of solution of $d=11$ supergravity on the sphere $\mathbb S^7$. It follows from
(\ref{51-39}) that this condition is not obligatory. Thus, if we choose
\begin{equation}\label{51-47}
F_{mnpq}=\pm\lambda S_{t[mn}S^{t}{}_{pq]}
\end{equation}
and take into account the conditions (\ref{51-40}), we get self-dual and anti--self-dual
solutions of Eq. (\ref{51-27}).
\par
Now we find restrictions that must be lay on the spaces $K$ and $M_4$. To this end, using
(\ref{51-28}) and (\ref{51-36}), we  rewrite Eq. (\ref{51-41}) in the form
\begin{equation}\label{51-49}
\overset\circ R{}^{i}_{jkl}=S^{i}_{jt}S^{t}_{kl}-hS^{i}_{jk;l},
\end{equation}
where $h=\pm1$. By summing on the indices $i$ and $k$, we obtained the Ricci tensor
\begin{equation}\label{51-42}
\overset\circ {R}_{mn}=6k^2g_{mn}.
\end{equation}
It follows from here that the $K$ is really an Einstein space. Substituting (\ref{51-42}) and
\begin{align}
F_{\mu\sigma\rho\lambda}F_{\nu}{}^{\sigma\rho\lambda}&=-6\rho^2g_{\mu\nu},\\
F_{mrpq}F_{n}{}^{rpq}&=24k^4\lambda^2g_{mn}
\end{align}
in the Einstein equation (\ref{51-11}), we get
\begin{equation}
\overset\circ {R}_{\mu\nu}=-10k^2g_{\mu\nu},\qquad 2\lambda^2=(12k)^{-2}.
\end{equation}
It follows from here that the four-dimensional spacetime $M_4$ is the anti--de Sitter space.
Note that all constants in the solutions are defined by the condition $\tilde e_{i}=ke_{i}$,
i.e. its values depend only on a selection of basis in the Malcev algebra $\mathbb M$.

\section{Conclusion}

In this paper, we have studied the Freund-Rubin-Englert mechanism of the $M_4\times K$
compactification of $d=11$ supergravity. We have shown that any threeform potential in 11
dimensions is given locally by the structure constants of a geodesic loop in an affinely
connected space. In particular, we have shown that any threeform potential with components
that take nonzero expectation values in $M_4$ is given by structure constants of the Lie
algebras (\ref{51-59}) and (\ref{51-69}). We have found the Englert's type solution of $d=11$
supergravity on the Einstein space $K$ and shown that the corresponding threeform potentials
are given locally by the Cayley structure constants. The solution is such that the affine
curvature tensor of $K$ is nonzero. It follows from here that the space $K$ may be not
parallelizable. Since everything considered in the paper is of a local character, the global
geometry of $K$ is not clear though.
\par
Note that the geodesic loops method that was used in the paper may be applied to the analysis
of M theory compactifications on singular manifolds with $G_2$ holonomy~\cite{atiy03, acha04}.
The point is that in addition to the compact Moufang loop $\mathbb S$ there exists a
noncompact nonassociative Moufang loop that is analytically isomorphic to the space ${\mathbb
S}^3\times {\mathbb R}^4$. This is exactly the asymptotically conical manifolds with $G_2$
holonomy.

\end{document}